# Detecting DNS Tunnels Using Character Frequency Analysis

Kenton Born
Kansas State University
kborn@ksu.edu

Dr. David Gustafson
Kansas State University
dag@ksu.edu

## Abstract

High-bandwidth covert channels pose significant risks to sensitive and proprietary information inside company networks. Domain Name System (DNS) tunnels provide a means to covertly infiltrate and exfiltrate large amounts of information passed network boundaries. This paper explores the possibility of detecting DNS tunnels by analyzing the unigram, bigram, and trigram character frequencies of domains in DNS queries and responses. It is empirically shown how domains follow Zipf's law in a similar pattern to natural languages, whereas tunneled traffic has more evenly distributed character frequencies. This approach allows tunnels to be detected across multiple domains, whereas previous methods typically concentrate on monitoring point to point systems. Anomalies are quickly discovered when tunneled traffic is compared to the character frequency fingerprint of legitimate domain traffic.

## Introduction

A common method of bypassing firewall restrictions is to tunnel restricted protocols inside legitimate protocols, encapsulating the blocked traffic and effectively hiding it from the firewall and network intrusion detection systems. Policy-restricted application layer protocols (ALP) can be transported inside HTTP or DNS traffic, defeating protocol-based traffic restrictions at network boundaries. Ozyman, TCP-over-DNS, Iodine, Dns2tcp, DNScat, and DeNiSe are a few of the many DNS tunneling applications available on the internet. While many methods have been proposed for detecting DNS tunnels, there are still no available tools that effectively prevent their infiltration and exfiltration capabilities.

DNS is a transactional protocol responsible for resolving domain names to IP addresses. Fully Qualified Domain Names (FQDN) are formed as a series of labels separated by periods. These labels break the domain into hierarchical subdomains where each subdomain is controlled by the next higher-level domain to the right. RFC 1035 (Mockapetris 1987) specifies that labels must be 63 octets or less, and the full domain must be 255 octets or less. The allowable characters are listed as a-Z, 0-9, and dashes. The top level domain (TLD) refers to the label furthest right in the domain name, while the Lowest Level Domain (LLD) refers to the label furthest left.

DNS is an ideal candidate for covert channels because it is poorly monitored and typically must be allowed passed network boundaries. Since DNS is transactional, it also provides sufficient capabilities for command and control channels into a protected network. However, DNS resolvers will often cache responses from a previously queried domain. This can be defeated by uniquely encoding exfiltrated data in the LLD, requiring the new domain to be resolved. When the query is received at the nameserver of an upper-level domain, it can decode the LLD and obtain the embedded data. The server is then free to send further commands or acknowledgements as a response to the query.

# Related Work

Many statistical approaches for detecting tunnels have been proposed that look at the captured flows of the traffic. *Web Tap* (Borders 2004) detects anomalies by looking at HTTP request regularity, inter-request delay, bandwidth usage, and transaction size. *Tunnel Hunter* (Crotti et al. 2007) approaches the problem from the IP layer by finding inconsistencies in inter-arrival time, order, and size of the packets. Focusing on a lower-layer protocol allows the tool to be applied more broadly against other types of tunnels. Crotti (Crotti et al. 2008) and Dusi (Dusi 2008) used *Tunnel Hunter* to fingerprint protocols and applications that were tunneled inside encrypted traffic. Despite being encrypted in SSH, the authors were able to fingerprint different protocol tunnels by analyzing their behavior at the IP layer. Although the above methods had significant success, a tunnel that mimics the pattern of legitimate traffic would deter these detection strategies.

Detecting DNS tunnels using anomaly detection to flag suspicious traffic was recently explored in the development of a tool known as *DnsTrap* (Hind 2009). In this tool, five attributes are used to train an Artificial Neural Network (ANN) to detect tunnels: the domain name, how many packets are sent to a particular domain, the average length of packets to that domain, the average number of distinct characters in the LLD, and the distance between LLD's. While the metrics chosen show promise, the use of neural networks is suspect and must be approached with caution because of their black -box nature and tendency to fall victim to generalization or over-fitting.

Many tools are available that collect DNS statistics and provide visualizations and analysis of the data. Two popular tools for DNS analysis are D*nstop (Dnstop 2009)* and D*sc* (Dsc 2009) . Dnstop builds display tables of common DNS statistics that could be of interest for both debugging and cyber security. Dsc, on the other hand, presents the user with various graphs that can be used to quickly analyze trends over time. Visualizing DNS traffic was examined deeper in (Ren et al. 2006), where the visual metaphor "*Flying Term*" was introduced. Several interesting visual representations of DNS traffic were given that rely on a human's spatial reasoning ability to quickly find anomalies that may be more difficult for a computer to detect. DNS analysis and visualization is also further explored in (Plonka and Barford 2008) using *TreeTop*. This tool classifies traffic based on context-aware clustering, differentiating between canonical, overloaded, and unwanted DNS queries. All of these works primarily focus on attacks against DNS, whereas the focus of this paper is on tunnels encapsulated in DNS. While many of the same strategies and themes could apply, these techniques have not been deeply explored for the purpose of DNS tunnel detection.

# Character Frequency Analysis

Character frequency analysis has been used many times when studying natural language. Initially examined by Zipf (Zipf 1932) and later by Shannon (Shannon 1951), it was shown that English has a distribution of characters that greatly affects the entropy (randomness) of the language. Patterns were found that made it easier to predict what the following bits of information would be as more information was collected. Since then, it has been used in several areas such as such as authorship identification and behavioral profiling. For example, Orebaugh (Orebaugh 2006) used character frequency analysis to profile users over instant messaging sessions. These profiles were then used to detect anomalous user activity.

Most forms of natural language have been shown to be biased toward certain characters and patterns. English, for example, is characterized by a disproportionately large number of characters from the popularized nonsense phrase *ETAOIN SHRDLU (*Relative frequencies of letters 2009). This phrase represents, from greatest to least, the order of frequencies of the most popular characters in the English language.

George Zipf (Zipf 1932) observed that the frequency of any word is inversely proportional to its rank in the frequency table (Zipf's Law). More specifically, a Zipfian distribution observes that the frequency of occurence of some event (P), as a function of rank (i), when the rank is determined by the frequency of occurence, is the power-law function $P_i \sim 1/i^a$ with the exponent a close to unity (Zipf's Law 2009).

While the Zipfian distribution is typically applied to natural languages, this work applies the concept to unigrams, bigrams, and trigrams in domain names. The motivation for this approach is that domain names should also have fingerprintable n-gram frequencies. While not all domain names are formed from the English language, nearly all natural languages have been shown to follow a Zipfian distribution, reducing their entropy. The most effective tunnels will compress, encrypt, then encode the data sent through the LLD of DNS queries. One measurement of a good encryption algorithm is how random the output of the encryption function appears (the greater the entropy, the harder it is to predict values from given ciphertext). My work attempts to exploit this property by showing how data embedded in the LLD of domains would not follow a Zipfian distribution similar to what is found in typical LLDs.

## Empirical Studies

Initial assumptions were first verified by comparing n-gram character frequencies in domains with their counterparts in the English language. This would determine whether domains had similar character patterns despite many domain names coming from a global pool of domains. The first test uses the 1,000,000 most popular domains (Top Sites 2009) with the LLDs and TLDs removed. These results are compared to the English language n-gram character frequencies provided in (Relative frequencies of letters 2009).

Below, figures 1a, 1b, and 1c compare the unigram, bigram, and trigram character frequencies to their English counterparts:

| English Unigrams | | Domain Unigrams | |
|---|---|---|---|
| LETTER | FREQUENCY | LETTER | FREQUENCY |
| e | 0.12702 | e | 0.10139 |
| t | 0.09056 | a | 0.08935 |
| a | 0.08167 | i | 0.07346 |
| o | 0.07507 | o | 0.07105 |
| i | 0.06966 | s | 0.06804 |
| n | 0.06749 | r | 0.06524 |
| s | 0.06327 | t | 0.06263 |
| h | 0.06094 | n | 0.06094 |
| r | 0.05987 | l | 0.04849 |
| d | 0.04253 | c | 0.03861 |
| l | 0.04025 | m | 0.03249 |
| c | 0.02758 | d | 0.03247 |
| u | 0.02758 | u | 0.03105 |
| m | 0.02406 | p | 0.02689 |

**Figure 1a: Comparison of English and domain unigram character frequencies**

| English Bigrams | | Domain Bigrams | |
| --- | --- | --- | --- |
| LETTER | FREQUENCY | LETTER | FREQUENCY |
| th | 0.03883 | in | 0.01702 |
| he | 0.03681 | er | 0.01550 |
| in | 0.02284 | an | 0.01333 |
| er | 0.02178 | re | 0.01290 |
| an | 0.02141 | es | 0.01271 |
| re | 0.01749 | ar | 0.01188 |
| nd | 0.01572 | on | 0.01135 |
| on | 0.01418 | or | 0.01051 |
| en | 0.01383 | te | 0.01017 |
| at | 0.01336 | al | 0.00976 |
| ou | 0.01286 | st | 0.00921 |
| ed | 0.01276 | ne | 0.00921 |
| ha | 0.01275 | en | 0.00897 |

**Figure 1b: Comparison of English and domain bigram character frequencies**

| English Trigrams | | Domain Trigrams | |
| --- | --- | --- | --- |
| LETTER | FREQUENCY | LETTER | FREQUENCY |
| the | 0.03508 | ing | 0.00498 |
| and | 0.01593 | ion | 0.00327 |
| ing | 0.01147 | ine | 0.00314 |
| her | 0.00822 | ter | 0.00314 |
| hat | 0.00651 | lin | 0.00306 |
| his | 0.00597 | ent | 0.00286 |
| tha | 0.00594 | the | 0.00285 |
| ere | 0.00561 | ers | 0.00258 |
| for | 0.00555 | and | 0.00240 |
| ent | 0.00531 | est | 0.00220 |
| ion | 0.00507 | tio | 0.00218 |
| ter | 0.00461 | tra | 0.00218 |
| was | 0.00461 | tor | 0.00212 |
| you | 0.00437 | art | 0.00204 |

**Figure 1c: Comparison of English and domain trigram character frequencies**

The first observation is that the unigram ranks in English have a very similar rank correlation to the unigram ranks of domains. Also, the changes in frequencies from one rank to the next are consistent between the two sets of data. For English, there was an average change in frequency between ranks of 0.059 per rank, while there was an average change of frequency of .057 per rank for domains. The biggest anomaly in the domain character ranks was the lack of the character 'h'. This letter is found often in English from words such as 'the', 'his', 'her', 'he', and 'she', which won't appear often in domain names.

While bigrams and trigrams did not match as well as unigrams, it must be taken into consideration that there is an exponentially growing pool of items in these cases. When the data is normalized over the size of the n-gram pool, digrams and trigrams actually show significant patterns. Similar to unigrams, "the" can be seen as an anomaly in bigrams and trigrams in domain names. Although change in frequency from one rank to the next was more significant in English at the bigram and trigram level, a consistent drop in frequency is still present for domains.

The following two graphs depict the similarities between English and domains when looking only at the change in frequency from one ranking to the next:

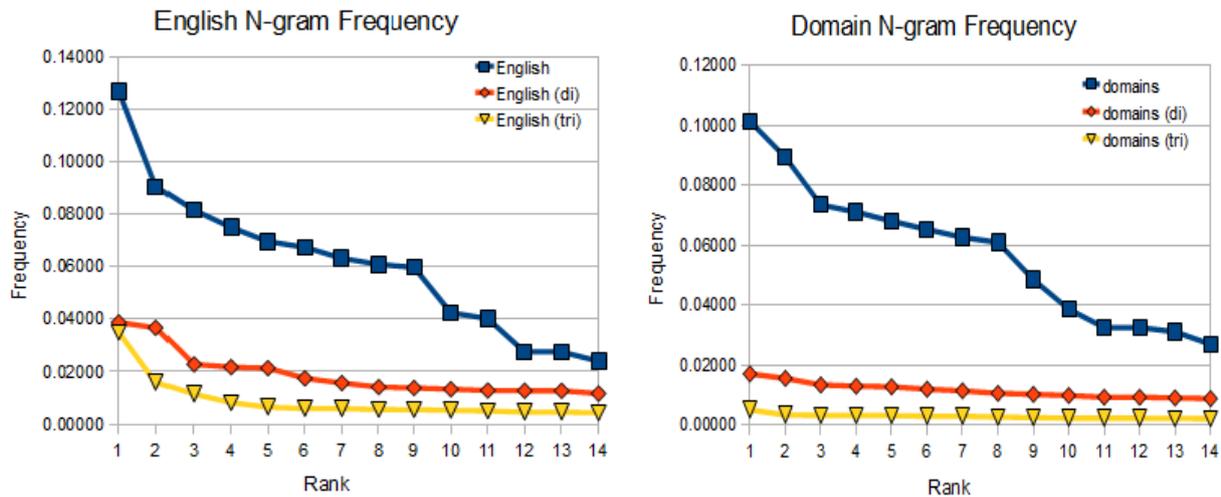

**Figure 2: English and domain n-gram frequencies by rank**

While the unigram drop in frequency appears much more pronounced than the others, it must be recognized that the frequency drop for bigrams and trigrams are distributed over an exponentially growing number of n-grams. Looking at the Zipfian dsitributionof n-gram frequencies appears to provide sufficient evidence for "language" regardless of the actual character rankings.

However, claims have been made that a Zipfian distribution is seen in many systems that are fed random data, explaining why it is seen in nearly all natural languages. Therefore, the next test compared the previous data with 1,000,000 new domain names, each composed of ten randomly generated characters. This test also served to mimic encrypted data in a tunnel, which should appear random if proper encryption was used. Below, the unigram ranks and frequencies of randomly generating domain names are shown next to the unigram frequencies of popular domain names:

| Random Domains | | Domain Unigrams | |
| --- | --- | --- | --- |
| LETTER | FREQUENCY | LETTER | FREQUENCY |
| i | 0.02722 | e | 0.10139 |
| 9 | 0.02715 | a | 0.08935 |
| z | 0.02710 | i | 0.07346 |
| c | 0.02710 | o | 0.07105 |
| f | 0.02709 | s | 0.06804 |
| 6 | 0.02708 | r | 0.06524 |
| o | 0.02707 | t | 0.06263 |
| k | 0.02707 | n | 0.06094 |
| q | 0.02707 | l | 0.04849 |
| j | 0.02706 | c | 0.03861 |
| 3 | 0.02705 | m | 0.03249 |
| 7 | 0.02705 | d | 0.03247 |
| 0 | 0.02704 | u | 0.03105 |
| 5 | 0.02406 | p | 0.02689 |

**Figure 3: Random domain unigram character frequencies for 1,000,000 domains**

Both unigram character ranks and the change in frequency from one rank to the next provide very strong evidence that character frequency analysis is a viable method of tunnel detection. Figure 3 above shows minimal correlation between the character ranks, and the change in frequency for random domains is only 1/20th of the change in frequency for popular domains.

For the above tests, a large amount of data was used to ensure accurate statistics. However, in an actual network, significantly less data can be used for the detection of tunnels. Without a large amount of data, it is possible that the the character ranks and frequency would be significantly skewed toward outliers in the data. Therefore, several of the previous tests were performed using only one hundred domains. Below, the results are compared to the previous results that used one million domains:

| 100 Domain Unigrams | | 1 Mil. Domain Unigrams | | 100 Domain Bigrams | | 1 Mil. Domain Bigrams | |
|---|---|---|---|---|---|---|---|
| LETTER | FREQUENCY | LETTER | FREQUENCY | LETTER | FREQUENCY | LETTER | FREQUENCY |
| e | 0.10496 | e | 0.10139 | in | 0.02048 | in | 0.01702 |
| o | 0.08455 | a | 0.08935 | er | 0.01707 | er | 0.01550 |
| i | 0.08163 | i | 0.07346 | li | 0.01707 | an | 0.01333 |
| a | 0.07872 | o | 0.07105 | re | 0.01707 | re | 0.01290 |
| r | 0.05977 | s | 0.06804 | ed | 0.01536 | es | 0.01271 |
| n | 0.05831 | r | 0.06524 | le | 0.01536 | ar | 0.01188 |
| t | 0.05831 | t | 0.06263 | ou | 0.01536 | on | 0.01135 |
| s | 0.04956 | n | 0.06094 | ar | 0.01195 | or | 0.01051 |
| d | 0.04665 | l | 0.04849 | es | 0.01195 | te | 0.01017 |
| l | 0.04373 | c | 0.03861 | me | 0.01195 | al | 0.00976 |
| u | 0.03644 | m | 0.03249 | or | 0.01195 | st | 0.00921 |
| b | 0.03207 | d | 0.03247 | so | 0.01195 | ne | 0.00921 |
| c | 0.03207 | u | 0.03105 | ba | 0.01024 | en | 0.00897 |
| g | 0.03061 | p | 0.02689 | di | 0.01024 | ra | 0.00875 |

**Figure 4: Unigram and bigram ranks and frequencies of 100 domains compared to 1,000,000 domains**

The results of the comparisons were very promising. Despite only using 100 domains, the n-gram character ranks and frequencies matched very well. The data collected over 100 domains had an average change of frequency of .055 per rank, comparing well with the .057 change in frequency for 1,000,000 domains. This approach was then used to compare 100 randomly generated domains with 1,000,000 popular domains:

| Random Domains | | Domain Unigrams | |
|---|---|---|---|
| LETTER | FREQUENCY | LETTER | FREQUENCY |
| z | 0.03800 | e | 0.10139 |
| y | 0.03500 | a | 0.08935 |
| 7 | 0.03400 | i | 0.07346 |
| q | 0.03400 | o | 0.07105 |
| 4 | 0.03400 | s | 0.06804 |
| w | 0.03400 | r | 0.06524 |
| 5 | 0.03300 | t | 0.06263 |
| 8 | 0.03100 | n | 0.06094 |
| p | 0.03000 | l | 0.04849 |
| - | 0.02900 | c | 0.03861 |
| 9 | 0.02900 | m | 0.03249 |
| m | 0.02900 | d | 0.03247 |
| v | 0.02900 | u | 0.03105 |
| x | 0.02800 | p | 0.02689 |

**Figure 5: Unigram ranks and frequencies of 100 random domains compared to 1,000,000 popular domains**

The unigram rank and frequencies of the random domains had minimal correlation with popular domains. An observation worth noting is that frequencies between ranks was more slightly more pronounced for 100 randoms domains then for 1,000,000 random domains. For 1,000,000 random domains, there was only a total difference in frequency of 0.00015 between the 1$^{st}$ and 10$^{th}$ ranked characters, while 100 random domains had a change in frequency of .0090 for the same ranks. This was due to outliers having a greater ability to skew the results when there is less data available. However, the change in frequency was still far less pronounced than the popular domain names, which had a total difference in frequency of .0628 between the 1$^{st}$ and 10$^{th}$ ranking characters. The Zipfian distribution of domain unigrams is shown below:

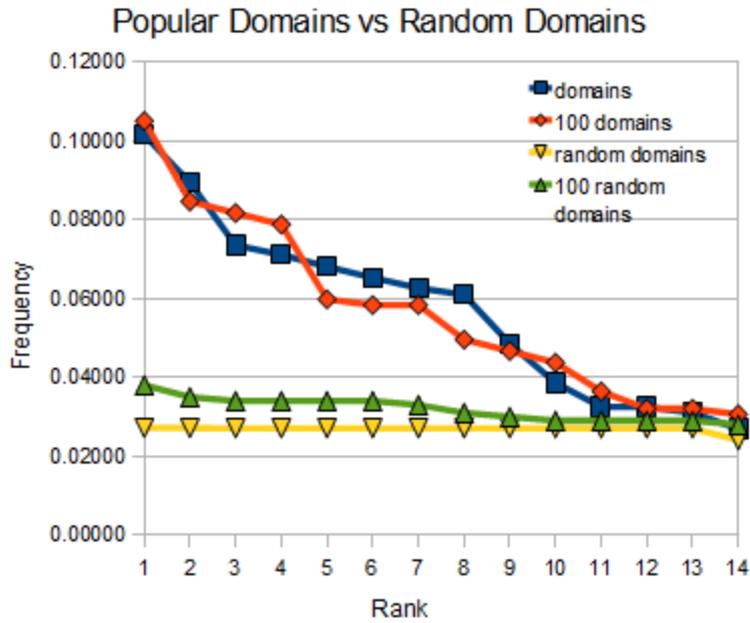

**Figure 6: Domain vs random domain unigram frequencies by rank**

With only 100 domains, both popular domains and randomly generated domains match very closely with the data from larger pools. It can also be seen that random domains will follow a Zipfian distribution less and less as the data set size increases. In contrast, popular domains will follow a Zipfian distribution more closely as the number of analyzed domains increases.

When analyzing DNS requests for tunnels, it becomes critical to concentrate on subdomains. To verify the validity of the approach for subdomains, unigram frequencies were once again calculated after stripping away the TLD and following label. Approximately 5000 DNS queries for subdomains were captured with Wireshark by creating a web crawler that visited sites from the previously used list of popular websites. The web crawler was allowed to span hosts, providing the most accurate depiction of traffic that would be seen from users surfing the internet in a networked environment. To obtain subdomains and nameservers queried by the ISP, a DNS response parser was created that could grab domains from all areas of the response, instead of simply relying on the DNS queries. Below, both subdomains and name server hosts are compared to the earlier formed fingerprint for domains:

| Subdomains | | Domain Unigrams | | NS Subdomains | | Domain Unigrams | |
| --- | --- | --- | --- | --- | --- | --- | --- |
| LETTER | FREQUENCY | LETTER | FREQUENCY | LETTER | FREQUENCY | LETTER | FREQUENCY |
| a | 0.08105 | e | 0.10139 | n | 0.20976 | e | 0.10139 |
| s | 0.08074 | a | 0.08935 | s | 0.19126 | a | 0.08935 |
| e | 0.06946 | i | 0.07346 | 1 | 0.06023 | i | 0.07346 |
| o | 0.06738 | o | 0.07105 | 2 | 0.06023 | o | 0.07105 |
| n | 0.06463 | s | 0.06804 | d | 0.04773 | s | 0.06804 |
| . | 0.05910 | r | 0.06524 | . | 0.04491 | r | 0.06524 |
| i | 0.05713 | t | 0.06263 | 0 | 0.03170 | t | 0.06263 |
| c | 0.04961 | n | 0.06094 | 3 | 0.02959 | n | 0.06094 |
| t | 0.04305 | l | 0.04849 | c | 0.02501 | l | 0.04849 |
| l | 0.04156 | c | 0.03861 | a | 0.02096 | c | 0.03861 |
| m | 0.03972 | m | 0.03249 | g | 0.02043 | m | 0.03249 |
| r | 0.03643 | d | 0.03247 | 4 | 0.02043 | d | 0.03247 |
| g | 0.03284 | u | 0.03105 | e | 0.01955 | u | 0.03105 |
| d | 0.02883 | p | 0.02689 | t | 0.01885 | p | 0.02689 |

**Figure 7: Unigram ranks and frequencies of subdomains and name server hosts compared to popular domains**

Subdomains are shown above to be very similar to domains in both character rank and frequency. It is interesting to note the increase in rank of the characters 'n' and 's'. This becomes more obvious when one looks strictly at the character frequency of name server hosts queried while the web crawler visited the domains. The subdomains from name servers are skewed by a heavy use of "ns#" patterns for name server hosts for domains such as "ns1.domain.com" and "ns2.domain.com".

While name server hosts do not match as strongly as other subdomains, it can be quickly seen that they have their own unique fingerprint that can be used for frequency analysis. While name server subdomains differ from typical subdomains, it is clearly seen that they still follow a fingerprintable pattern that can be used for comparisons. Below, a Zipfian distribution can be seen in subdomains (with a noticeable anomaly for the 'n' and 's' in name server subdomains):

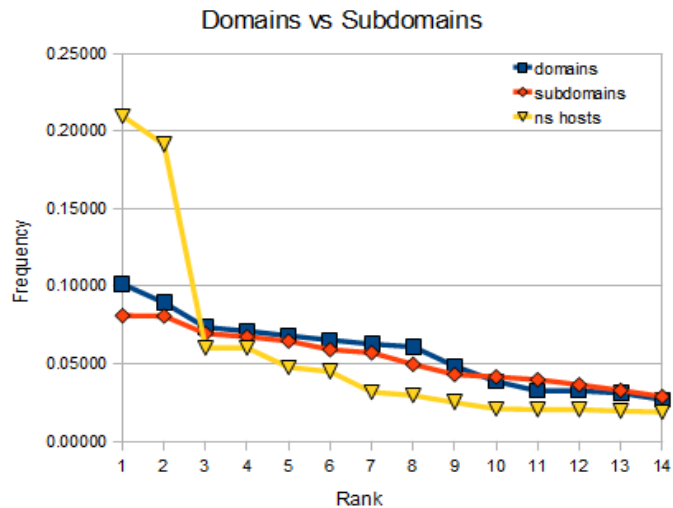

Figure 9: Subdomain unigram frequencies by rank

To verify the legitimacy of character frequency analysis, the above approach was used to collect statistics on several popular DNS tunnels. Sample traffic from Iodine, Dns2tcp, and TCP-Over-DNS was collected. In order to simulate data being exfiltrated from a network, all three programs were used to tunnel an SCP session of a pdf file from one host to another. One-hundred consecutive DNS queries were analyzed for each tunnel. Below, the character ranks and frequencies are compared to the results provided earlier for subdomains:

| Iodine | | Subdomains | |
| --- | --- | --- | --- |
| LETTER | FREQUENCY | LETTER | FREQUENCY |
| a | 0.04969 | a | 0.08105 |
| c | 0.03360 | s | 0.08074 |
| s | 0.03329 | e | 0.06946 |
| b | 0.03217 | o | 0.06738 |
| q | 0.03186 | n | 0.06463 |
| l | 0.03177 | . | 0.05910 |
| o | 0.03132 | i | 0.05713 |
| n | 0.03119 | c | 0.04961 |
| m | 0.03096 | t | 0.04305 |
| t | 0.03038 | l | 0.04156 |
| w | 0.03033 | m | 0.03972 |
| f | 0.03020 | r | 0.03643 |
| r | 0.03020 | g | 0.03284 |
| g | 0.03006 | d | 0.02883 |

Figure 10a: Unigram ranks and frequencies of 100 Iodine packets compared to subdomains

| Dns2tcp | | Subdomains | |
|---|---|---|---|
| LETTER | FREQUENCY | LETTER | FREQUENCY |
| n | 0.04515 | a | 0.08105 |
| k | 0.03794 | s | 0.08074 |
| c | 0.03770 | e | 0.06946 |
| r | 0.03569 | o | 0.06738 |
| b | 0,03470 | n | 0.06463 |
| u | 0.03137 | . | 0.05910 |
| t | 0.03121 | i | 0.05713 |
| d | 0.03105 | c | 0.04961 |
| m | 0.03097 | t | 0.04305 |
| s | 0.03097 | l | 0.04156 |
| x | 0.03097 | m | 0.03972 |
| p | 0.03080 | r | 0.03643 |
| a | 0.03048 | g | 0.03284 |
| o | 0.02918 | d | 0.02883 |

**Figure 10b: Unigram ranks and frequencies of 100 Dns2tcp packets compared to subdomains**

| TCP-Over-DNS | | Subdomains | |
|---|---|---|---|
| LETTER | FREQUENCY | LETTER | FREQUENCY |
| k | 0.04177 | a | 0.08105 |
| g | 0.04021 | s | 0.08074 |
| b | 0.03731 | e | 0.06946 |
| j | 0.03726 | o | 0.06738 |
| m | 0.03531 | n | 0.06463 |
| o | 0.03497 | . | 0.05910 |
| d | 0.03475 | i | 0.05713 |
| e | 0.03462 | c | 0.04961 |
| c | 0.03414 | t | 0.04305 |
| v | 0.03406 | l | 0.04156 |
| n | 0.03401 | m | 0.03972 |
| q | 0.03393 | r | 0.03643 |
| a | 0.03358 | g | 0.03284 |
| l | 0.03358 | d | 0.02883 |

**Figure 10c: Unigram ranks and frequencies of 100 TCP-Over-DNS packets compared to subdomains**

When looking at character ranks, all three tunnels managed to match subdomains significantly better than the randomly generated domains from before, but still fell below the consistency seen in legitimate traffic. Most of the tunnels used base64 encoding while the randomly generated domains only used 37 characters (no upper-cased letters) at maximum entropy (randomness). Since these tunnels used both lowercase and uppercase letters to increase entropy (but were counted equally for the totals above), the "number" characters fell to lower ranks in the tunnels.

While the character ranks did not prove as useful in this case, the change in frequency between ranks was very telling. Below, the change in character frequency in subdomains is contrasted with the change in character frequency in the three tunnels:

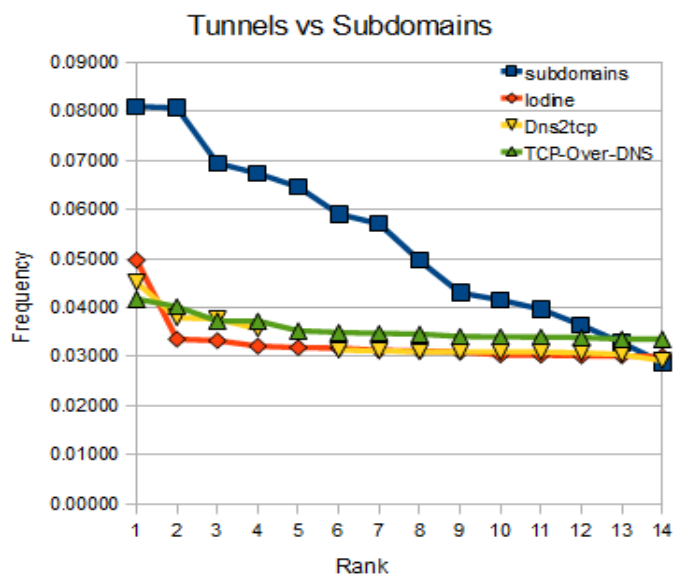

**Figure 11: Tunnel unigram frequencies by rank for 100 queries vs subdomains**

The above graph shows a clear delineation between legitimate traffic and tunneled data. While many of the tunnels have a significantly increased frequency in the first two ranks, the rest of the ranks match closely with the randomly generated domains from before. The increased frequency in the first two ranks can be attributed to session id's and counters that had little change from one DNS request to the next.

## Conclusions and Future Work

It has been empirically shown how DNS tunnels may be detected by analyzing the character frequencies of DNS queries and responses. Both domains and subdomains were shown to match the Zipfian distribution found in the English language. Fingerprints were developed for domains, subdomains, and name server hosts that can be used for detecting anomalous traffic being tunneled through DNS. This method may be used to detect tunnels that span multiple domains instead of focusing on point to point systems.

Future work will involve building a tool that combines visualization and quantitative analysis for DNS tunnel detection. Empirical studies with this tool will provide sufficient data for creating a program for real-time DNS traffic monitoring and analysis. Character frequency analysis will be combined with multiple forms of tunnel detection to significantly reduce the effective bandwidth of covert channels. Future work will also determine the amount of padding necessary to mitigate detection through character frequency analysis.